\documentclass[12pt]{article}

\usepackage{epsfig}
\usepackage{amssymb}
\usepackage{graphics,graphpap}

\usepackage{amsmath}
\usepackage[letterpaper]{hyperref}

\renewcommand{\thefootnote}{\fnsymbol{footnote}}

\begin{document}

%%%%%%%%%% Title page
\begin{titlepage}
\begin{flushright}
\begin{tabular}{l}
IPPP/08/29\\
DCPT/08/58
\end{tabular}
\end{flushright}
\vskip1.5cm
\begin{center}
{\Large \bf \boldmath
Constraining New Physics in $B^0\to \pi^+ \pi^-$ with\\[5pt]
Reparametrization Invariance and QCD Factorization}
\vskip1.3cm 
{\sc
Patricia Ball\footnote{Patricia.Ball@durham.ac.uk}
and
Aoife Bharucha\footnote{a.k.m.bharucha@durham.ac.uk}
} \vskip0.5cm
{\em IPPP, Department of Physics,
University of Durham, Durham DH1 3LE, UK}

\vskip1.5cm

%{\em Version of \today}

\vskip1.5cm

{\large\bf Abstract\\[10pt]} \parbox[t]{\textwidth}{
Usually, $B^0\to\pi^+\pi^-$ decays are expressed in terms of weak amplitudes explicitly dependent on the CKM weak phase $\alpha$ or $\gamma$. In this letter, we show that the  weak amplitudes can be rewritten such that a manifest dependence on $\beta$ emerges instead. Based on this, we constrain new-physics contributions to the CP-violating phase $\phi_d$ in $B^0$--$\bar{B}^0$ mixing. Further, we apply reparametrization invariance and use QCD factorization predictions to investigate the bounds on an additional new-physics amplitude in $B^0\to\pi^+\pi^-$. 
}

\end{center}
\end{titlepage}

\setcounter{footnote}{0}
\renewcommand{\thefootnote}{\arabic{footnote}}
\renewcommand{\theequation}{\arabic{equation}}

\newpage

%%%%%%%%%%%%%%%%% Main text

One of the greatest successes of the $B$ factories BaBar and Belle  is the precise determination of the CP-violating phase $\phi_d$ in $B$ mixing \cite{HFAG}. In the 
Standard Model (SM), and using the Wolfenstein parametrization of the CKM matrix, $\phi_d$ is related to $\beta$, one of the angles of the unitarity triangle, as $\phi_d=2\beta$.
As $B$ mixing is a loop process, the experimentally determined angle $\phi_d$ might in fact not equal $2\beta$, but be polluted by the effects of new-physics (NP) particles propagating in loops, thereby contributing an additional CP violating phase, see for instance  Ref.~\cite{BF}.\footnote{NP at tree level this is highly disfavoured as its impact on $B$ mixing would exceed the observed effects by far -- unless the mass scales involved are in the $\sim 10\,$TeV range, see Ref.~\cite{treeBmixing}.} It is therefore of considerable interest to study any methods by which one can constrain an additional NP contribution to $\phi_d$. In this letter we shall show that the process $B^0\to\pi^+\pi^-$ can be used to this effect.

The set of neutral and charged $B \to \pi\pi$ decays has been extensively studied as a means of determining the angle $\alpha$ (or $\gamma$) of the unitarity triangle. The lack of a theoretically clean calculation of the strong amplitudes and phases involved can be overcome by exploiting isospin symmetry, see Ref.~\cite{Gronau:1990ka}, commonly referred to as the Gronau-London method. It involves relating the various experimental observables (branching ratios and CP asymmetries) in $B\to \pi\pi$ to extract both the hadronic amplitudes determining these decays and the weak phase $\alpha$.  As an alternative to isospin, and in order to avoid $B^0\to\pi^0\pi^0$ decays, the use of U-spin has been explored in Refs.~\cite{Fleischer:1999pa} to extract $\gamma$ from $B^0 \to \pi^+\pi^-$ and the U-spin related decay $B_s\to K^+ K^-$. In a conceptionally different approach the relevant strong amplitudes are calculated (as opposed to extracted from experiment), using QCD factorization (QCDF) \cite{QCDF,BS,QCDFNNLO,Bell09} or effective field theory methods (SCET) \cite{SCET}. 
The advantage here is that less experimental input is needed, the disadvantage that the calculation is performed in a limit of QCD where the $b$ quark is assumed to be very heavy.
In any case, all these analyses put the emphasis on constraining the angles $\gamma$ or $\alpha$.

In this letter, we show that it is possible to express the decay amplitude in terms of $\phi_d$ and $\beta$, without any explicit reference to the angles $\alpha$ or $\gamma$. In the SM, the resulting expression allows the extraction of the relevant hadronic parameters from $B^0\to\pi^+\pi^-$ data alone, which can be compared to the theoretical calculation in QCDF. Beyond the SM, we  study the possible presence of NP in this decay, which might contribute through $B^0$--$\bar{B}^0$ mixing via a NP contribution to $\phi_d$  or through an additional NP amplitude. 

We begin with a reminder of the parametrization used to extract $\gamma$. The amplitude for $\bar{B}^0 \to \pi^+\pi^-$ is given in the form: 
\begin{equation}
\mathcal{A}(\bar{B}^0 \to \pi^+\pi^-)=\lambda_c A_c+\lambda_u A_u\,,
\label{eq:cconv}
\end{equation} 
where $\lambda_q=V^*_{qd}V^{\phantom{*}}_{qb}$, and $A_c$, $A_u$ are strong amplitudes. $A_u$ is dominated by tree diagrams, whereas the only contributions to $A_c$ are from penguin diagrams.  
The corresponding time-dependent CP asymmetry is given by:
\begin{eqnarray}
A_{\pm}(t)&=&\frac{\Gamma(B^0(t) \to \pi^+\pi^-)-\Gamma(\bar{B}^0(t) \to \pi^+\pi^-)}{\Gamma(B^0(t) \to \pi^+\pi^-)+\Gamma(\bar{B}^0(t) \to \pi^+\pi^-)}
\\\nonumber
&=&C_{\pm}\cos(\Delta m t)-S_{\pm}\sin(\Delta m t)\,.\label{eq:ACPt}
\end{eqnarray}
The experimental observables $C_{\pm}$ and $S_{\pm}$ can be expressed in terms of $\lambda_{\pm}$:
\begin{equation}
S_{\pm} = \frac{2\, {\rm Im}(\lambda_{\pm})}{1+|\lambda_{\pm}|^2}\,,\qquad
C_{\pm} = \frac{1-|\lambda_{\pm}|^2}{1+|\lambda_{\pm}|^2}\,,\label{eq:SC}
\end{equation}
where $\lambda_{\pm}$ is given by
\begin{equation}
\lambda_{\pm}=
e^{- i \phi_d} \,\frac{\mathcal{A}(\bar{B}^0 \to \pi^+\pi^-)}{\mathcal{A}(B^0 \to \pi^+\pi^-)}\,.
\end{equation}
Parametrizing the amplitudes as in Eq.~(\ref{eq:cconv}), we have
\begin{equation} 
\lambda_\pm=e^{-i\phi_d}\,\frac{e^{-i\gamma}-re^{i\delta}}{e^{i\gamma}-re^{i\delta}}
\end{equation}
with $r e^{i\delta}=A_c/(A_uR_b)$, $\gamma={\rm\arg}(-\lambda_c/\lambda_u)$ and \begin{equation}\label{Rb}
R_b=\left|\frac{\lambda_u}{\lambda_c}\right| = \frac{1-\lambda^2/2}{\lambda}\,\frac{|V_{ub}|}{|V_{cb}|}\,.
\end{equation}
Numerical values for these and other CKM-related quantities are collected in Tab.~\ref{tab:0}.
\begin{table}
$$
\renewcommand{\arraystretch}{1.2}
\addtolength{\arraycolsep}{3pt}
\begin{array}{|c||c|l|}
\hline
\mbox{Parameter} & \mbox{Value} & \mbox{Source}\\
\hline
\hline
\lambda & 0.2257^{+0.0009}_{-0.0010} & \mbox{PDG~}\cite{PDG}\\
|V_{cb}| & (41.2\pm 1.1)\times 10^{-3} & \mbox{PDG~}\cite{PDG}\\
|V_{ub}| & (3.93\pm 0.36)\times 10^{-3} & \mbox{PDG~}\cite{PDG}\\
\beta_{b\to ccs} & (21.1\pm 0.9)^\circ & \mbox{HFAG~}\cite{HFAG}\\
\beta{\mbox{\tiny tree}} & (23.9\pm 3.3)^\circ & \mbox{this paper, Eq.~(\ref{betatree})}\\
\gamma & (77^{+30}_{-32})^\circ &  \mbox{PDG~}\cite{PDG}\\
R_b & 0.412\pm 0.039 & \mbox{this paper, Eq.~(\ref{Rb})}\\
\left|\frac{V_{td}}{V_{ts}}\right| & 0.214\pm 0.005 & \mbox{\cite{VtdVtslatt}}\\
R_t & 0.928\pm 0.024 & \mbox{this paper, Eq.~(\ref{Rt})}\\
\hline
\end{array}
\addtolength{\arraycolsep}{-3pt}
$$
\caption[]{\small CKM parameters used in this letter.}\label{tab:0}
\end{table}
The observables $S_\pm$ and $C_\pm$ are given by 
\begin{eqnarray}
S_{\pm} &=& −\frac{\sin (\phi_d + 2\gamma) - 2r \sin (\phi_d +\gamma) \cos\delta + r^2 \sin \phi_d}{1 - 2r \cos \gamma \cos\delta + r^2}\,, \label{eq:SCgammaA}\\
C_{\pm} &=&-\frac{2r \sin\gamma \sin\delta}{1 - 2r \cos\gamma \cos\delta + r^2}
\,.
\label{eq:SCgammaB}
\end{eqnarray}
In the absence of penguin contributions, $r=0$ and the determination of $\phi_d+2\gamma$ would be completely analogous to that of $\phi_d$ from $B^0\to J/\psi K_S$. Realistically, $r$ is expected to be a small, but non-zero number, which makes the extraction of $\gamma$ more challenging. 

We now show how a different parametrization of the decay amplitude (\ref{eq:cconv}) replaces the explicit dependence of $\lambda_\pm$ on $\gamma$ by one on $\beta$. Using $\beta={\rm arg}(-\lambda_t/\lambda_c)$, one can trade the dependence on $\gamma$ for one on $\beta$ by exploiting the unitarity of the CKM matrix and exchanging $\lambda_u$ for $-\lambda_c-\lambda_t$: 
\begin{eqnarray}
\mathcal{A}(\bar{B}^0 \to \pi^+\pi^-)&=&\lambda_c B_c+\lambda_t B_t\nonumber\\
 &=&\lambda_c(B_c-R_t e^{i \beta} B_t)\,,\label{eq:Abeta}
\end{eqnarray}
where $B_c=A_c-A_u$, $B_t=-A_u$ and 
\begin{equation}\label{Rt}
R_t=\left|\frac{\lambda_t}{\lambda_c}\right| = \frac{1}{\lambda}\, \frac{|V_{td}|}{|V_{ts}|} \left\{ 1 - \frac{1}{2}\left( 1 - 2 R_b \cos\gamma\right) \lambda^2 + O(\lambda^4)\right\}.
\end{equation}
Note that $B_c$ and $B_t$ are both dominated by tree-level decays as they both contain $A_u$.

With this parametrization of the decay amplitude, $\lambda_{\pm}$ becomes
\begin{eqnarray}
\lambda_{\pm}&=&e^{- i \phi_d} \left(\frac{1-R_t R_{tc} e^{i\beta}}{1-R_t R_{tc} e^{-i\beta}}\right)\\
&=&e^{-i \phi_d} \left(\frac{1-d e^{i\theta_d} e^{i\beta}}{1-d e^{i\theta_d} e^{-i\beta}}\right),
\end{eqnarray}
where $R_{tc} = B_t/B_c$ and  $d=|R_t R_{tc}|$, $\theta_d=\arg( R_t R_{tc})$. Note that unlike $r$, $d$ is not suppressed, but expected to be of order $1$ (as $R_t$ is also close to 1). The CP-violating observables in (\ref{eq:ACPt}) now read 
\begin{eqnarray}
S_{\pm}&=&\frac{d^2\sin(2\beta-\phi_d)+2d\cos\theta_d \sin(\phi_d-\beta)-\sin(\phi_d)}{d^2-2d \cos\beta\cos\theta_d+1}\,, \label{eq:SCbetaA} \\
C_{\pm}&=&-\frac{2d \sin\beta\sin\theta_d}{d^2-2d\cos\beta\cos\theta_d+1}\,.
\label{eq:SCbetaB}
\end{eqnarray}
Obviously (\ref{eq:SCbetaA}), (\ref{eq:SCbetaB}) are not independent of (\ref{eq:SCgammaA}), (\ref{eq:SCgammaB}), but related by the unitarity constraint
\begin{equation}\label{15}
R_te^{i\beta}+R_be^{-i\gamma}-1=0\,.
\end{equation}
The advantage of expressing $S_{\pm}$ and $C_{\pm}$ in terms of $\beta$ instead of $\gamma$ is that, at least in the SM, there is now only one manifest weak phase. This implies that, with $R_t$ determined from $B$ mixing, both $d$ and $\theta_d$ can be extracted from experiment and compared to theoretical calculations, for example QCDF. This is independent of any information from the decay $B^0\to\pi^0\pi^0$ whose branching ratio continues to be difficult to understand in the framework of QCDF or SCET. 

The most accurate measurement of $\phi_d$ is via mixing in $B^0$ decays to CP eigenstates of charmonium. The CP asymmetry averaged over these channels provides a direct measurement of $\sin\phi_d=0.673\pm0.023$, so that in the SM $\beta=(21.1\pm0.9)^\circ$ \cite{HFAG}\footnote{There is an ambiguity in this result, as $\beta=(68.9\pm1.0)^\circ$ is also a solution. However, this is excluded at the 95\% confidence level by a Dalitz plot analysis of $B^0\to\bar{D}^0h^0$ where $h^0=\pi^0$, $\omega$, $\eta$ \cite{Dalitz},  and by a time-dependent angular analysis of $B^0\to J/\psi K^{*0}$ \cite{timedep}.}. It is also possible to derive $\beta$ from tree-process measurements only, based on $\gamma$ and $|V_{ub}|$. Taking $\gamma$ and $|V_{ub}|$ from Ref.~\cite{PDG}, see Tab.~\ref{tab:0}, we find $\beta_{\rm tree}$ using 
\begin{equation}\label{betatree}
\sin\beta_{\rm tree} = \frac{R_b\sin\gamma}{\sqrt{1-2 R_b \cos\gamma + R_b^2}}\,,\qquad 
\cos\beta_{\rm tree} = \frac{1-R_b\cos\gamma}{\sqrt{1-2 R_b \cos\gamma + R_b^2}}\,,
\end{equation}
which results in $\beta_{\rm tree}=(23.9^{+3.4}_{-3.2})^\circ$ (in the following analysis we use $\beta_{\rm tree}=(23.9\pm3.3)^\circ$). Both values of $\beta$ are compatible with each other, but we will use the latter one to obtain constraints on a NP contribution to $\phi_d$.
\begin{table}
\begin{center}
\renewcommand{\arraystretch}{1.2}
\addtolength{\arraycolsep}{3pt}
\begin{tabular}{|l||c|c|}
\hline
Experiment & $S_{\pm}$& $C_{\pm}$\\
\hline
\hline
BaBar \cite{Aubert:2008sb}&$-0.68 \pm 0.10 \pm 0.03$&$-0.25 \pm 0.08 \pm 0.02$\\
Belle \cite{Ishino:2006if}&$-0.61 \pm 0.10 \pm 0.04$&$-0.55 \pm 0.08 \pm 0.05$\\
\hline
HFAG \cite{HFAG}&$-0.65 \pm 0.07$&$-0.38 \pm 0.06$\\
\hline
\end{tabular}
\addtolength{\arraycolsep}{-3pt}
\end{center}
\caption[]{\small Experimental results for $S_{\pm}$, $C_{\pm}$ from BaBar and Belle and the HFAG average.}\label{tab:1}
\vspace*{10pt}
\begin{center}
\renewcommand{\arraystretch}{1.2}
\addtolength{\arraycolsep}{3pt}
\begin{tabular}{|l||c|c||c|c|}
\hline
&\multicolumn{2}{c||}{$\beta_{b\to ccs}=(21.1\pm 0.9)^\circ$}&\multicolumn{2}{c|}{$\beta_{\mbox{\tiny tree}}=(23.9\pm3.3)^\circ$}\\
\cline{2-5}
& $d$& $\theta_d$ & $d$& $\theta_d$\\
\hline
\hline
BaBar&$0.790\pm 0.031$&$0.068 \pm 0.025$&$0.775\pm 0.037$&$0.075 \pm 0.028$\\
Belle&$0.803 \pm 0.033$&$0.158 \pm 0.041$&$0.789 \pm 0.040$&$0.174 \pm 0.049$\\
\hline
HFAG&$0.796\pm 0.021$&$0.104 \pm 0.020$&$0.782\pm 0.027$&$0.115 \pm 0.025$\\
\hline
QCDF&$0.825^{+0.034}_{-0.052}$&$-0.021^{+0.043}_{-0.042}$&$0.825^{+0.034}_{-0.052}$&$-0.021^{+0.043}_{-0.042}$\\
\hline
\end{tabular}
\addtolength{\arraycolsep}{-3pt}
\end{center}
\caption[]{\small Comparison of $d$ and $\theta_d$ derived from experimental results to QCDF predictions for $\beta_{b\to ccs}$ from $b\to ccs$ transitions and $\beta_{\mbox{\tiny tree}}$ derived from tree decays.}\label{tab:2}
\end{table}

From the experimental data collected in Tab.~\ref{tab:1}, we find the values of $d$ and $\theta_d$ given in Tab.~\ref{tab:2}. The high quality of the experimental results leads to small uncertainties on $d$, typically 5\%, and moderate uncertainties on $\theta_d$. As expected, $d$ is of order 1, while $\theta_d$ is rather small.\footnote{Actually, there is a discrete ambiguity in the determination of $d$ and $\theta_d$ which yields a second solution $d\sim 0.3$ and $\theta_d\sim 1$. We discard this solution as $d$ is the ratio of two tree-dominated amplitudes and hence expected to be close to 1; $d\sim 0.3$ would imply a massive NP amplitude which can not be generated in any NP models we are aware of.} 

How do these results compare to theoretical predictions?
As mentioned earlier, QCDF provides us with a framework for computing the individual amplitudes contributing to $B^0\to\pi^+ \pi^-$, and the relative phases between them. Complete results for both tree and penguin contributions are available to NLO accuracy \cite{BS}. The result for  $A_c/A_u$ is:\footnote{Recently the NNLO calculation of the tree amplitude has been
completed in Ref.~\cite{Bell09}. We do not use this result in our
analysis, as we also require the penguin amplitude at NNLO.
Nevertheless, we have checked numerically that using this value 
would alter $A_c/A_u$ by 4 to 5\%.}
\begin{equation}\label{17}
\frac{A_c}{A_u}=-0.122^{+0.033}_{-0.063}-0.024^{+0.047}_{-0.048}i\,.
\end{equation}
Therefore using $B_c=A_c-A_u$, $B_t=-A_u$, we find 
\begin{eqnarray}
R_{tc}&=&\frac{B_t}{B_c} = 0.891^{+0.026}_{-0.050}-0.019^{+0.037}_{-0.038}i\,.
\end{eqnarray}
As $R_t$ is, by definition, a positive number, $\theta_d$ is given by the phase of $R_{tc}$, as seen in Tab.~\ref{tab:2}. The theoretical prediction is ca.\ 3$\sigma$ away from the experimental (HFAG) result and has the opposite sign.
From Eq.~(\ref{eq:SCbetaB}) it is clear that $\theta_d$ is mainly constrained by $C_\pm$. While Tab.~\ref{tab:1} clearly shows that the results from BaBar and Belle are not completely in agreement, they {\em do} agree on the sign and require a positive $\theta_d$. 
This situation is similar to that for $B\to K\pi$ decays, where the sign of the observed direct CP violation is difficult to reconcile with QCDF predictions. The source of the discrepancy is independent of any NP contributions to $\Delta m_{d,s}$, from which $R_t$ is determined, as $\theta_d$ is independent of $R_t$. It is also unrelated to any NP contributions to the $B$ mixing phase $\phi_d$, as using the tree-level determination of $\beta$, $\beta_{\rm tree}$, for the explicit $\beta$ dependence renders the discrepancy even slightly worse (right-hand side of Tab.~\ref{tab:2}). An explanation in terms of NP could be that an additional NP amplitude contributes to $B^0\to\pi^+\pi^-$, and we explore this possibility below. Otherwise, the discrepancy may be related to neglected higher order terms in QCDF -- either from radiative corrections or terms suppressed by inverse powers of the $b$ quark mass. 
As for extreme values of the QCDF input parameters a positive $\theta_d$ {\em is} possible, it would be interesting to see what choice of these parameters yields positive $\theta_d$ and whether this choice agrees with the scenarios advocated in Refs.~\cite{BS,QCDFNNLO,Bell09} to reconcile the QCDF predictions for $B(B^0\to\pi^0\pi^0)$ with experimental data.

\begin{figure}
\begin{center}
\begin{tabular}{cc}
\includegraphics[scale=.75]{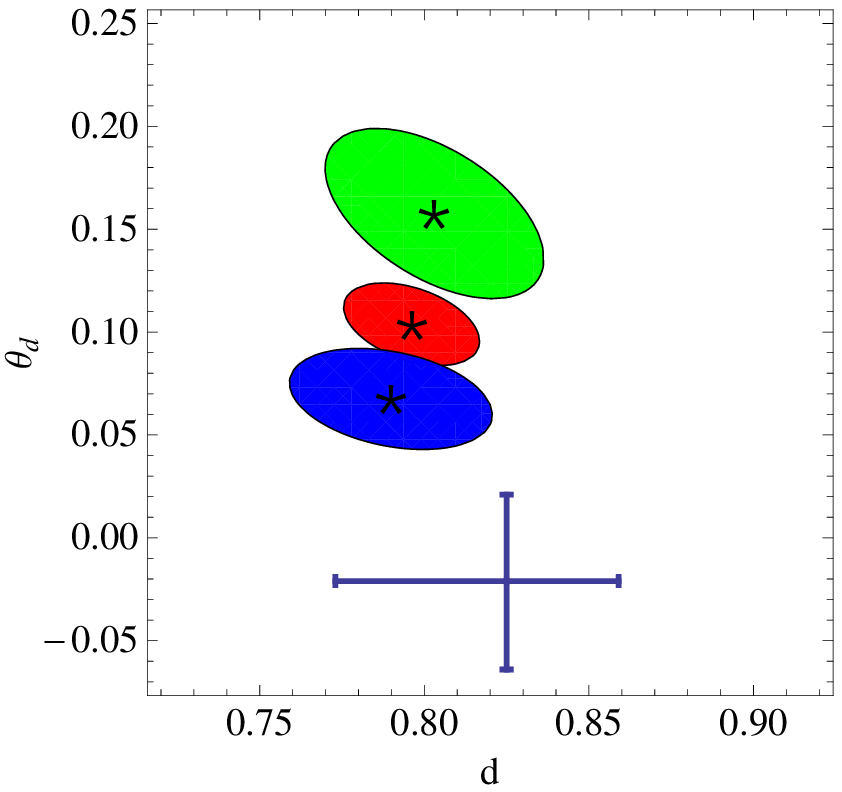}&\includegraphics[scale=.75]{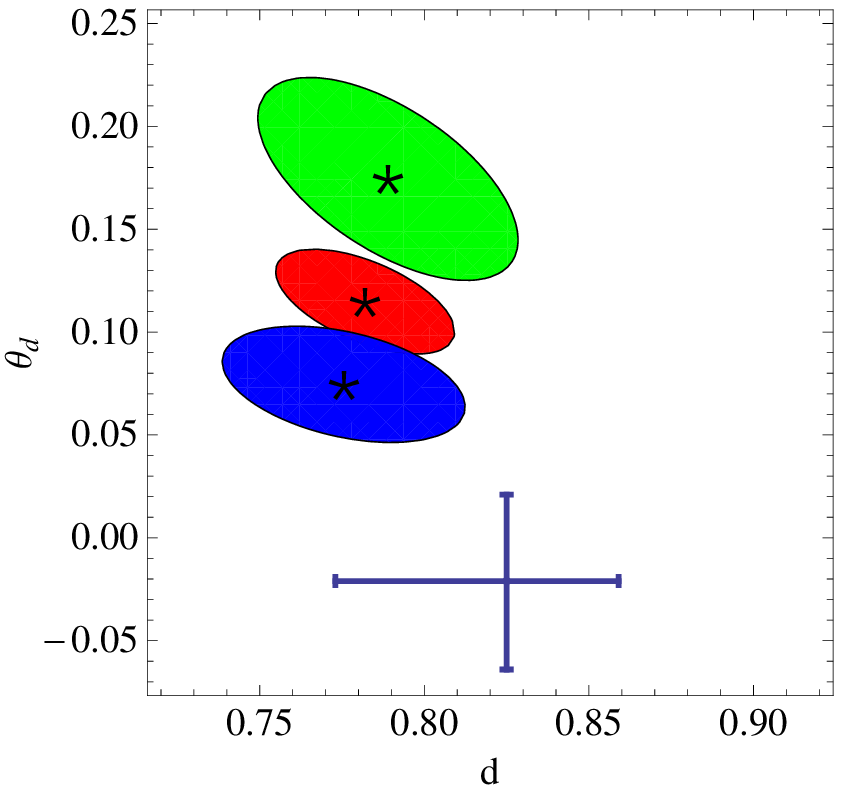}
\end{tabular}
\caption[]{\small The allowed experimental region for $d$ and $\theta_d$ are plotted for the case $\beta=(21.1\pm 0.9)^\circ$ on the left and $\beta=(23.9\pm3.3)^\circ$ on the right. The blue (lower) region corresponds to the results for $S_\pm$ and $C_\pm$ of BaBar, the green (upper) region corresponds to those of Belle and the red (central) region corresponds to the average given by HFAG as in Tab.~\ref{tab:1}. The blue point with error bars represents the QCDF prediction.}
\label{fig:1}
\end{center}
\end{figure}

In order to calculate $d$ from QCDF we also need $R_t$. This can be determined from the ratio of the mass differences in the $B_s$ and $B_d$ systems. CDF has made a very clean measurement of $\Delta m_s$  \cite{Abulencia:2006mq}, which can be turned into a value for $R_t$ using lattice information on the relevant hadronic parameters \cite{VtdVtslatt} and Eq.~(\ref{Rt}). The result is  given in Tab.~\ref{tab:0}. This value of $R_t$ allows for a limited contribution of NP to the mass differences in the $B_{d,s}$ systems: from the unitarity relation (\ref{15}), using $R_t$ and $\beta$ as input, one finds $R_b\approx 0.36$ and $\gamma\approx 68^\circ$, in good agreement with the values quoted in Tab.~\ref{tab:0}.\footnote{The agreement is not perfect for $R_b$. If, however, one uses the result for $|V_{ub}|\approx 3.5\times 10^{-3}$ from exlusive decays, instead of the larger PDG value, $R_b\approx 0.36$ and the agreement is near perfect.} Combining the QCDF calculation with this value for $R_t$, we find that the QCDF prediction agrees with the experimental results within 1$\sigma$. The experimental results and QCDF predictions for $d$ and $\theta_d$ are shown in Fig.~\ref{fig:1}. The analogous QCDF predictions needed in the ``standard'' parametrization of $S_\pm$, Eq.~(\ref{eq:SCgammaA}), and $C_\pm$, Eq.~(\ref{eq:SCgammaB}), can be found by dividing (\ref{17}) by $R_b$:
\begin{equation}
 r=0.30^{+0.15}_{-0.09},\qquad \delta+\pi=0.194^{+0.384}_{-0.382}\,.
\end{equation}
Note the marked difference in the relative errors for ($d$, $\theta_d$) and ($r$, $\delta$), which is an important advantage to the parametrization (\ref{eq:Abeta}) of the $B^0\to\pi^+\pi^-$ amplitude and will reduce the uncertainties in the constraints on a NP amplitude contribution to this decay.

Let us now discuss constraints on $\phi_d-2\beta$, based on the QCDF results in Tab.~\ref{tab:2}. Assuming there is no NP amplitude contributing to the decay, one can obtain $\beta$ using $\phi_d$ and $d$, $\theta_d$ from QCDF  as input. This results in $\beta = 19.9^\circ$ with a minimum $\chi^2$ of 63. Leaving $\theta_d$ as fit parameter instead, we find $\beta = (16.4\pm 6.5)^\circ$ which shows that the above discrepancy in $\theta_d$ from QCDF and from experiment is largely irrelevant for NP contributions to $\phi_d$.  
Both values are consistent with $\phi_d/2=(21.1\pm 0.9)^\circ$, and and also with $\beta=(23.9\pm3.3)^\circ$ determined from $\gamma$ and $|V_{ub}|$. However, it is still interesting to note the two results for $\phi_d/2-\beta$ are of opposite sign. If there was indeed a NP contribution to $\phi_d$, we would expect to see common trend in the sign of $\phi_d/2-\beta$ from various determination. As this does not seem to be the case, we take this as an indication that any NP contribution to $\phi_d$ is indeed small. 

\begin{figure}
\begin{center}
\begin{tabular}{cc}
\includegraphics[scale=0.84]{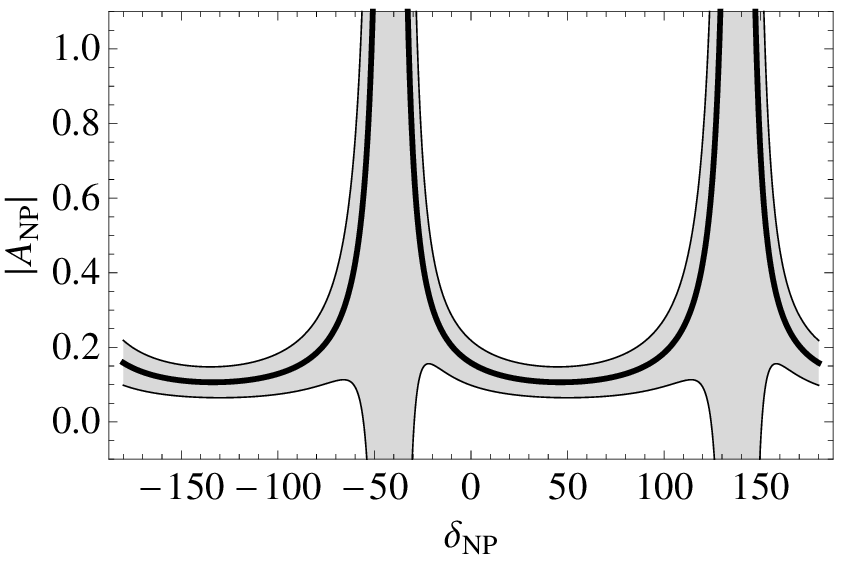}&\includegraphics[scale=0.84]{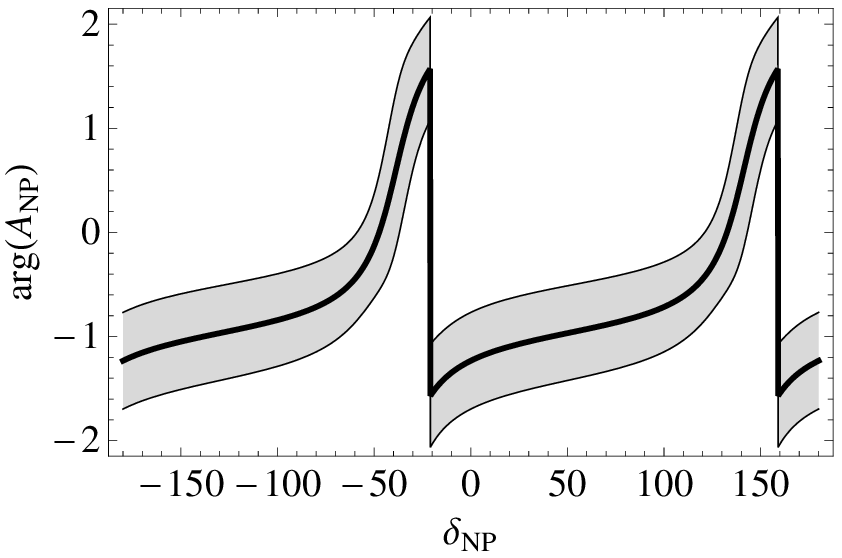}\\
\end{tabular}
\caption[]{\small $|A_{\rm NP}|$ and $\arg(A_{\rm NP})$ plotted against the weak phase of the new physics amplitude, $\delta_{\rm NP}$. The thick black lines represent the results found using the experimental average value of $\sin(\phi_d)$, the grey regions represent the uncertainty.}
\label{fig:2}
\end{center}
\end{figure}

Given this result, the decay $B^0\to\pi^+\pi^-$ seems then a very suitable place to constrain any NP contributions to the decay amplitude. Such contributions are possible in a variety of NP models, e.g.\ the 2-Higgs doublet model \cite{2Higgs} or R-parity violating SUSY \cite{Rparity}. 
We therefore introduce the (complex) NP amplitude $A_{\rm NP}$ and a new weak phase $\delta_{\rm NP}$, such that 
\begin{equation}
\lambda_{\pm}=e^{-i \phi_d} \left(\frac{1-d e^{i\theta_d} e^{i\beta}+A_{\rm NP} e^{i\delta_{\rm NP}}}{1-d e^{i\theta_d} e^{-i\beta}+A_{\rm NP}e^{-i\delta_{\rm NP}}}\right).
\end{equation}
Here $d$ and $\theta_d$ are SM quantities, given by the QCDF values stated earlier, in Tab.~\ref{tab:2}.  The above expression for $\lambda_\pm$ is actually not suitable to constrain $A_{\rm NP}$ and $\delta_{\rm NP}$: in Ref.~\cite{Botella:2005ks}, it was shown that a given amplitude, with an arbitrary number of distinct weak phases, can always be expressed in terms of any two weak phases. Using this so-called reparametrization invariance, $\lambda_{\pm}$ can be expressed in terms of the two weak phases $\phi_d$ and $\beta$:\footnote{Eq.~(\ref{x}) was already obtained in 
Ref.~\cite{Botella:2005ks}, as (48).}
\begin{equation}\label{x}
\lambda_{\pm}=e^{-i \phi_d} \left(\frac{1-d' e^{i\theta_d'} e^{i\beta}}{1-d' e^{i\theta_d'} e^{-i \beta}}\right),
\end{equation}
where
\begin{equation}
d' e^{i\theta_d'}=\frac{d e^{i\theta_d}-A_{\rm NP}\sin(\delta_{\rm NP})/\sin\beta}{1+A_{\rm NP} \sin(\beta- \delta_{\rm NP})/\sin\beta}\,.\label{eq:Anp}
\end{equation}
Note that $d' e^{i\theta_d'}\to d' e^{i\theta_d'}$ under a CP transformation, so $\theta_d'$ is indeed a strong phase. In Eq.~(\ref{x}), $\phi_d$ is, by definition, the phase measured in $B$ mixing, i.e.\ from $b\to ccs$ transitions, while $\beta$ is obtained from tree-level processes, i.e.\ $\phi_d=(42.2\pm 1.8)^\circ$ and $\beta=(23.9\pm3.3)^\circ$.
The values for $d'$ and $\theta_d'$ are obviously identical to the experimental results for $d$ and $\theta_d$ in Tab.~\ref{tab:2}. The resulting constraints on $A_{\rm NP}$ and $\delta_{\rm NP}$ are shown in Fig.~\ref{fig:2}. Depending on the value of $\delta_{\rm NP}$, large NP contributions $|A_{\rm NP}|$ are possible. Note that these results are mainly due to the discrepancy between $\theta_d'$ from experiment and $\theta_d$ from QCDF. In particular, even for $\delta_{\rm NP}=0$ a non-zero $|A_{\rm NP}|\approx 0.1$ is needed. Also note that the allowed valued for $|A_{\rm NP}|$ can be as large, or even larger, than the SM (QCDF) prediction for $d$. Although it would be interesting to interpret the NP amplitude in terms of the NP models mentioned, making use of constraints from other processes, this is beyond the scope of this letter. 

In summary, we have investigated the effect of NP on the decay $B^0\to \pi^+\pi^-$. We first analyzed the CP asymmetries in the SM, using a particularly convenient parametrization of the decay amplitude which depends on only one weak phase, $\beta$. We compared the experimental results of the relevant hadronic quantities with those from theoretical calculations, using  QCDF. We found that while the size of the strong amplitude is consistent, its predicted phase $\theta_d$ deviates from experiment by $\sim 3\sigma$. It would be interesting to study whether the choice of QCDF input parameters minimizing this discrepancy is the same needed to reduce that between the predicted and observed branching ratio of $B^0\to\pi^0\pi^0$. We then analyzed the implications for NP contributions to the decay. Under the assumption that NP modifies only the phase in $B$ mixing, the discrepancy in the strong phase was largely irrelevant and the NP contributions to the mixing phase found to be consistent with 0. We then allowed for a NP amplitude contributing to the decay amplitude of $B^0\to \pi^+\pi^-$ and analyzed the CP asymmetries exploiting the constraints imposed by reparametrization invariance. Here we found scope for large NP contributions, which are driven by the discrepancy between $\theta_d$ from experiment and from QCDF. An analysis of this NP amplitude within specific models is beyond the scope of this letter.


\begin{thebibliography}{99}

\bibitem{HFAG}
  E.~Barberio {\it et al.}  [Heavy Flavor Averaging Group],
  %``Averages of b-hadron and c-hadron Properties at the End of 2007,''
  arXiv:0808.1297 [hep-ex], and online updates at www.slac.stanford.edu/xorg/hfag.
  %%CITATION = ARXIV:0808.1297;%%

\bibitem{BF}
A.J.~Buras, R.~Fleischer, S.~Recksiegel and F.~Schwab,
  %``New aspects of B $\to$ pi pi, pi K and their implications for rare
  %decays,''
  Eur.\ Phys.\ J.~{\bf C45} (2006) 701
  [arXiv:hep-ph/0512032];\\
  %%CITATION = HEP-PH 0512032;%%
M.~Bona {\it et al.}\  [UTfit Collaboration],
  %``The UTfit collaboration report on the status of the unitarity triangle
  %beyond the standard model. I: Model-independent analysis and minimal flavour
  %violation,''
  JHEP {\bf 0603} (2006) 080
  [arXiv:hep-ph/0509219];\\
  %%CITATION = HEP-PH 0509219;%%
P.~Ball and R.~Fleischer,
  %``Probing new physics through $B$ mixing: Status, benchmarks and prospects,''
  Eur.\ Phys.\ J.\  C {\bf 48} (2006) 413
  [arXiv:hep-ph/0604249].
  %%CITATION = EPHJA,C48,413;%%

\bibitem{treeBmixing}
P.~Ball, J.~M.~Frere and J.~Matias,
  %``Anatomy of Mixing-Induced CP Asymmetries in Left-Right-Symmetric Models
  %with Spontaneous CP Violation,''
  Nucl.\ Phys.\  B {\bf 572} (2000) 3
  [arXiv:hep-ph/9910211];\\
  %%CITATION = NUPHA,B572,3;%%
M.~Bona {\it et al.}  [UTfit Collaboration],
  %``Model-independent constraints on $\Delta$ F=2 operators and the scale of
  %new physics,''
  JHEP {\bf 0803} (2008) 049
  [arXiv:0707.0636 [hep-ph]].
  %%CITATION = JHEPA,0803,049;%%

\bibitem{Gronau:1990ka}
  M.~Gronau and D.~London,
  %``Isospin analysis of CP asymmetries in B decays,''
  Phys.\ Rev.\ Lett.\  {\bf 65} (1990) 3381.
  %%CITATION = PRLTA,65,3381;%%

\bibitem{Fleischer:1999pa}
  R.~Fleischer,
  %``New strategies to extract beta and gamma from B/d --> pi+ pi- and  B/s -->
  %K+ K-,''
  Phys.\ Lett.\  B {\bf 459} (1999) 306
  [arXiv:hep-ph/9903456];\\
  %%CITATION = PHLTA,B459,306;%%
  A.~Soni and D.~A.~Suprun,
  %``Determination of gamma from charmless B+ --> M0 M+ decays using U-spin,''
  Phys.\ Lett.\  B {\bf 635} (2006) 330
  [arXiv:hep-ph/0511012].
  %%CITATION = PHLTA,B635,330;%%

%\bibitem{Silva:1993sv}
%  J.~P.~Silva and L.~Wolfenstein,
%  %``Determining the penguin effect on CP violation in B0 $\to$ pi+ pi-,''
%  Phys.\ Rev.\  D {\bf 49} (1994) 1151
%  [arXiv:hep-ph/9309283].
%  %%CITATION = PHRVA,D49,1151;%%

\bibitem{QCDF}
  M.~Beneke, G.~Buchalla, M.~Neubert and C.~T.~Sachrajda,
  %``{QCD} factorization for B --> pi pi decays: Strong phases and CP  violation
  %in the heavy quark limit,''
  Phys.\ Rev.\ Lett.\  {\bf 83} (1999) 1914
  [arXiv:hep-ph/9905312].
  %%CITATION = PRLTA,83,1914;%%

\bibitem{BS}
M.~Beneke and S.~J\"ager,
  %``Spectator scattering at NLO in non-leptonic B decays: Tree amplitudes,''
  Nucl.\ Phys.\  B {\bf 751} (2006) 160
  [arXiv:hep-ph/0512351];
  %%CITATION = NUPHA,B751,160;%%
%M.~Beneke and S.~Jager,
  %``Spectator scattering at NLO in non-leptonic B decays: Leading penguin
  %amplitudes,''
  Nucl.\ Phys.\  B {\bf 768} (2007) 51
  [arXiv:hep-ph/0610322].
  %%CITATION = NUPHA,B768,51;%%

\bibitem{QCDFNNLO}
V.~Pilipp,
  %``Hard spectator interactions in B to pi pi at order alpha_s^2,''
  Nucl.\ Phys.\  B {\bf 794} (2008) 154
  [arXiv:0709.3214 [hep-ph]];\\
  %%CITATION = NUPHA,B794,154;%%
G.~Bell,
  %``NNLO Vertex Corrections in charmless hadronic B decays: Imaginary part,''
  Nucl.\ Phys.\  B {\bf 795} (2008) 1
  [arXiv:0705.3127 [hep-ph]].
  %%CITATION = NUPHA,B795,1;%%

\bibitem{Bell09}
G.~Bell,
  %``NNLO vertex corrections in charmless hadronic B decays: Real part,''
  arXiv:0902.1915 [hep-ph].
  %%CITATION = ARXIV:0902.1915;%%

\bibitem{SCET}
C.~W.~Bauer, I.~Z.~Rothstein and I.~W.~Stewart,
  %``SCET analysis of B --> K pi, B --> K anti-K, and B --> pi pi decays,''
  Phys.\ Rev.\  D {\bf 74} (2006) 034010
  [arXiv:hep-ph/0510241].
  %%CITATION = PHRVA,D74,034010;%%

\bibitem{PDG}
  C.~Amsler {\it et al.}  [Particle Data Group],
  %``Review of particle physics,''
  Phys.\ Lett.\  B {\bf 667} (2008) 1.
  %%CITATION = PHLTA,B667,1;%%

\bibitem{VtdVtslatt}
  E.~Gamiz {\it et al.} [HPQCD Collaboration],
  %``Neutral $B$ Meson Mixing in Unquenched Lattice QCD,''
  arXiv:0902.1815 [hep-lat].
  %%CITATION = ARXIV:0902.1815;%%

\bibitem{Dalitz}
  B.~Aubert {\it et al.}  [BABAR Collaboration],
  %``Measurement of $\cos$ 2 beta in $B^0 \to D^{(*)}$ h0 Decays with a
  %Time-Dependent Dalitz Plot Analysis of $D \to K^0_{S} \pi^{+} \pi^{-}$,''
  Phys.\ Rev.\ Lett.\  {\bf 99} (2007) 231802
  [arXiv:0708.1544 [hep-ex]];\\
  %%CITATION = PRLTA,99,231802;%%
  K.~Abe {\it et al.},
  %``Measurement of phi(1) using anti-B0 --> D (K0(S) pi+ pi-) h0,''
  arXiv:hep-ex/0507065.
  %%CITATION = HEP-EX/0507065;%%

\bibitem{timedep}
  B.~Aubert {\it et al.}  [BABAR Collaboration],
  %``Ambiguity-free measurement of $\cos(2\beta)$: Time-integrated and
  %time-dependent angular analyses of $B \to J/\psi K \pi$,''
  Phys.\ Rev.\  D {\bf 71} (2005) 032005
  [arXiv:hep-ex/0411016];\\
  %%CITATION = PHRVA,D71,032005;%%
  R.~Itoh {\it et al.}  [Belle Collaboration],
  %``Studies of CP violation in B --> J/psi K* decays,''
  Phys.\ Rev.\ Lett.\  {\bf 95} (2005) 091601
  [arXiv:hep-ex/0504030].
  %%CITATION = PRLTA,95,091601;%%
\bibitem{Aubert:2008sb}
  B.~Aubert {\it et al.}  [BABAR Collaboration],
  %``Measurement of CP Asymmetries and Branching Fractions in B0 -> pi+ pi-, B0
  %-> K+ pi-, B0 -> pi0 pi0, B0 -> K0 pi0 and Isospin Analysis of B -> pi pi
  %Decays,''
  arXiv:0807.4226 [hep-ex].
  %%CITATION = ARXIV:0807.4226;%%

\bibitem{Ishino:2006if}
  H.~Ishino {\it et al.}  [Belle Collaboration],
  %``Observation of Direct CP-Violation in B0 ---> pi+ pi- Decays and
  %Model-Independent Constraints on phi2,''
  Phys.\ Rev.\ Lett.\  {\bf 98} (2007) 211801
  [arXiv:hep-ex/0608035].
  %%CITATION = PRLTA,98,211801;%%

\bibitem{Abulencia:2006mq}
  A.~Abulencia {\it et al.}  [CDF],
  %``Measurement of the $B^0_{s}-\bar{B}^0_s$ Oscillation Frequency,''
  Phys.\ Rev.\ Lett.\  {\bf 97} (2006) 062003
  [arXiv:hep-ex/0606027].
  %%CITATION = PRLTA,97,062003;%%

%\bibitem{Buchalla:2004tw}
%  G.~Buchalla and A.~S.~Safir,
%  %``CP violation in B --> pi+ pi- and the unitarity triangle,''
%  Eur.\ Phys.\ J.\  C {\bf 45} (2006) 109
%  [arXiv:hep-ph/0406016].
%  %%CITATION = EPHJA,C45,109;%%

\bibitem{2Higgs}
  Z.~j.~Xiao, C.~S.~Li and K.~T.~Chao,
  %``Charmless hadronic decays $B \to$ PP, PV, VV and new physics effects in the
  %general two Higgs doublet models,''
  Phys.\ Rev.\  D {\bf 63} (2001) 074005
  [arXiv:hep-ph/0010326].
  %%CITATION = PHRVA,D63,074005;%%

\bibitem{Rparity}
  G.~Bhattacharyya, A.~Datta and A.~Kundu,
  %``$R$-Parity violation in $B \to \pi^+ \pi^-$ decay,''
  J.\ Phys.\ G {\bf 30} (2004) 1947
  [arXiv:hep-ph/0212059].
  %%CITATION = JPHGB,G30,1947;%%

\bibitem{Botella:2005ks}
  F.~J.~Botella and J.~P.~Silva,
  %``Reparametrization invariance of B decay amplitudes and implications for
  %new physics searches in B decays,''
  Phys.\ Rev.\  D {\bf 71} (2005) 094008
  [arXiv:hep-ph/0503136].
  %%CITATION = PHRVA,D71,094008;%%

\end{thebibliography}
\end{document}